\documentclass[11pt]{article}
\usepackage{Blois,epsfig,amssymb}

\def\be{\begin{equation}}
\def\ee{\end{equation}}
\def\bea{\begin{eqnarray}}
\def\eea{\end{eqnarray}}

\newcommand{\beq}{\begin{equation}}
\newcommand{\eq}{\end{equation}}
\newcommand{\rb}{\underline{r}}
\newcommand{\kb}{\underline{k}}

\newcommand{\zb}{\bar{z}}

\newcommand{\intfeyn}{\int\limits_0^1}

\begin{document}
\begin {flushright}
CPHT-PC 056.0905

LPT-ORSAY 05-64

\end{flushright}
\vspace*{2cm}

\begin{center}
\Large{\textbf{XIth International Conference on\\ Elastic and  
Diffractive Scattering\\ Ch\^{a}teau de Blois, France, May 15 - 20,  
2005}}
\end{center}
\vspace*{2cm}

\title{BORN ORDER STUDY  OF $\gamma^*\gamma^*\to \rho \rho$ AT VERY HIGH ENERGY}

\author{ B. PIRE$^1$,
  L. SZYMANOWSKI$^{2,3}$ and S. WALLON$^4 $}

\address{${}^1$\,CPHT~\footnote{Unit{\'e} mixte C7644 du CNRS}, \'Ecole
Polytechnique, 91128 Palaiseau, France \\[0.2\baselineskip]
${}^2$\,Soltan Institute for Nuclear Studies, Warsaw, Poland
\\[0.2\baselineskip]
${}^3$\,Universit\'e  de Li\`ege,  B4000  Li\`ege,  
Belgium\\[0.2\baselineskip]
${}^4$\,LPT~\footnote{Unit{\'e} mixte 8627 du CNRS}, Universit\'e  
Paris-Sud, 91405-Orsay, France  \\
}

\maketitle\abstracts{
We calculate the cross-section for the diffractive exclusive process
$\gamma^*_L (Q_1^2)\gamma^*_L(Q_2^2) \to \rho^0_L \rho^0_L,$ in view of its  
study in the future high energy
  $e^+e^--$ linear collider.
The Born order approximation of the amplitude  is completely calculable  
in the hard region
  $Q_1^2 \,, Q_2^2 \gg \Lambda^2_{QCD}.$
The resulting cross-section is large enough for this process to
be measurable with foreseen luminosity and energy, for $Q_1^2$ and  
$Q_2^2$ in the range of a
few $GeV^2.$ }

\section{Motivation}
High energy QCD dynamics may be tested by 
the next generation of $e^+e^--$colliders.
Exclusive processes such as 
double vector meson production
by two highly virtual  
photons give access to the kinematical regime in which the  
perturbative
approach is justified. If additionally one selects the events with
comparable photon virtualities, the perturbative Regge dynamics of QCD
  of the BFKL \cite{bfkl} type should dominate with respect to the
conventional partonic evolution of DGLAP  type.
We propose \cite{PSW} to study the electroproduction
of two  longitudinally polarized $\rho-$mesons,
\beq
\label{process}
\gamma^*_L(q_1)\;\gamma^*_L(q_2) \to \rho^0_L(k_1)  \;\rho^0_L(k_2)\,,
\ee
for arbitrary values of $t=(q_1-k_1)^2,$ with $s \gg -t.$
The Born
approximation estimate of the
cross section proves the feasibility of a dedicated experiment.
The extension of this study by taking into  
account of BFKL
   evolution is given in \cite{EPSW}. The case of 
transverse photon polarizations will be 
studied soon. At lower  
energy, some experimental data exist
   for $Q_2^2$ small  \cite{L3} which may be
analysed \cite{apt} in terms of generalized distribution amplitudes.

\section{Impact representation}
\label{impact}

The impact representation of the scattering amplitude  has the form
\be
\label{M}
{\cal M} = is\;\int\;\frac{d^2\,\kb}{(2\pi)^4\kb^2\,(\rb -\kb)^2}
{\cal J}^{\gamma^*_L(q_1) \to \rho^0_L(k_1)}(\kb,\rb -\kb)\;
{\cal J}^{\gamma^*_L(q_2) \to \rho^0_L(k_2)}(-\kb,-\rb +\kb)\,,
\ee
where ${\cal J}^{\gamma^*_L(q_1) \to \rho^0_L(k_1)}(\kb,\rb -\kb)$
(${\cal J}^{\gamma^*_L(q_2) \to \rho^0_L(k_2)}(\kb,\rb -\kb)$)
are the impact factors corresponding  to the
transition of
$\gamma^*_L(q_1)\to \rho^0_L(k_1)$ ($\gamma^*_L(q_2)\to \rho^0_L(k_2)$)  
via the
$t-$channel exchange of two gluons.
The amplitude (\ref{M}) depends linearly on $s$, since these impact  
factors
are $s$-independent.
Calculations of the impact factors in the Born approximation are  
standard; they read, using the notation $\bar z = 1-z$,
\bea
\label{if}
{\cal J}^{\gamma^*_L(q_i) \to \rho_L(k_i)}(\kb,\rb -\kb)
= \int\limits_0^1 dz_i z_i \, \zb_1 \, \phi(z_1)8 \pi^2 \alpha_s
\frac{e}{\sqrt{2}} \frac{\delta^{a b}}{2 N_c} Q_i \, f_\rho 
\rm{P_P(z_i,\kb,\rb,\mu_i)}\,,
\eea
where
\beq
\label{defPP}
\rm{P_P(z_i,\kb,\rb,\mu_i)}=
  \frac{1}{z_i^2\rb^2 + \mu_i^2} +
\frac{1}{\zb_i^2\rb^2 + \mu_i^2}  
  - \frac{1}{(z_i\rb -\kb)^2 + \mu_i^2} - \frac{1}{(\zb_i\rb
-\kb)^2
+ \mu_i^2}
\eq
originates from the impact factor of quark pair production from a
longitudinally polarized photon, with two t-channel exchanged gluons.
In formula (\ref{defPP}) the collinear approximation has
been made, namely the relative 
transverse momenta of quarks with respect to the mesons 
momenta have been neglected in the hard part of the amplitude. We denote
$\mu_i^2=Q_i^2 \; z_i \; \zb_i $
in the case of massless quarks.

In Eq.(\ref{if}), $\phi$ is the distribution
amplitude of the produced longitudinally polarized
$\rho-$mesons. For simplicity, we use the asymptotic distribution  
amplitude
\beq
\phi(z)=6 \, z \, (1-z)\,.
\eq
The only remaining part of the quark phase space after applying the
collinear approximation are the  
integration with respect
to quark longitudinal fractions of meson momenta $z_1$ and $z_2.$

Combining Eqs.(\ref{M}, \ref{if}, \ref{defPP}), the
amplitude can be expressed
as
\beq
\label{MCalgeneral}
{\cal M} = i\, s\, 2 \,\pi \,\frac{N_c^2-1}{N_c^2}\, \alpha_s^2  
\,\alpha_{em} \, f_\rho^2\, Q_1\, Q_2 \, \intfeyn d z_1 \, d  
z_2 \, z_1  
\, \zb_1 \, \phi(z_1)\, z_2
\, \zb_2 \, \phi(z_2) {\rm M}(z_1,\, z_2)\,,
\eq
with
\bea
\label{defM}
{\rm M}(z_1,\, z_2)=\int \frac{d^2 \kb}{\kb^2 (\rb-\kb)^2} &&\left[  
\frac{1}{z_1^2\rb^2 + \mu_1^2} +
\frac{1}{\zb_1^2\rb^2 + \mu_1^2}
  - \frac{1}{(z_1\rb -\kb)^2 + \mu_1^2} - \frac{1}{(\zb_1\rb
-\kb)^2
+ \mu_1^2} \right] \nonumber \\
&&\hspace{-3cm} \times \left[ \frac{1}{z_2^2\rb^2 + \mu_2^2} +
\frac{1}{\zb_2^2\rb^2 + \mu_2^2}
  - \frac{1}{(z_2\rb -\kb)^2 + \mu_2^2} - \frac{1}{(\zb_2\rb
-\kb)^2
+ \mu_2^2} \right]\,.
\eea 

Note that at moderate values of $s,$ apart from diagrams with gluon
exchange
in $t-$channel considered here, one should also take into account of diagrams 
with quark exchange. We do not consider them here since we restrict ourselves
to the asymptotical region of large $s.$ 

\section{Forward cross section}
In the simpler case $t= t_{min}$ ({\it i.e.}
  $\rb=0$), the integral over $\kb$ can readily be performed and gives
\beq
\label{Mlimite}  
M(z_1,z_2) = \frac{4 \pi}{z_1 \, \zb_1 z_2 \, \zb_2 Q_1^2 \, Q_2^2\,
(z_1 \, \zb_1 Q_1^2 - z_2 \, \zb_2 Q_2^2)} \ln \frac{z_1 \, \zb_1  
Q_1^2}{z_2 \, \zb_2 Q_2^2}\,.
\eq
The amplitude ${\cal M}$ given by Eq.(\ref{MCalgeneral}) can then be  
computed
analytically from Eq.(\ref{Mlimite}) through double integration over   
$z_1$ and $z_2$ :
\bea
\label{resultMmin} 
&&\hspace{-.7cm}{\cal M}_{t_{min}}= -i s \, \frac{N_c^2-1}{N_c^2} \,  
\alpha_s^2 \, \alpha_{em} \,
 f_\rho^2 \, \frac{9 \pi^2}{2} \,  
\frac{1}{Q_1^2
   Q_2^2}  \left[6 \, \left(R + \frac{1}{R}\right) \ln^2 R + \,12\,
\left(R-\frac{1}{R}\right) \ln R \nonumber \right.\\
&&\hspace{-.8cm}\left.+\, 12 \,  \left(R + \frac{1}{R}\right)\,+\,
\left(3 \,R^2 +\, 2\,+\frac{3}{R^2}\right) \left(\,\ln(1-R)\,\ln^2 R
-\ln(R+1)\,\ln^2 R -2 \,{\rm Li_2}(-R)\, \ln R \nonumber \right. \right.\\
&&+ \left. \left. \, 2 \,{\rm Li_2}(R)\, \ln R \,+\,2\,{\rm Li_3}(-R)\,-\,2\,  
{\rm Li_3}(R)\right)\right]\,,
\eea
where $R=Q_1/Q_2.$

In the special case where $Q=Q_1=Q_2,$ it simplifies  to
\beq
\label{amplitudeR1}
{\cal M}_{t_{\min}}(Q_1=Q_2) \sim \, -i \,s \,  \frac{N_c^2-1}{N_c^2} \,  
\alpha_s^2 \, \alpha_{em} \,
 f_\rho^2 \, \frac{9 \pi^2}{2 Q^4}\, (24 -  
28 \, \zeta(3))\,.
\eq

\begin{figure}[htb]
\vspace{-1cm}
\begin{minipage}[t]{105mm}
\epsfxsize=7.7cm{\centerline{\hspace{-2.75cm}\epsfbox{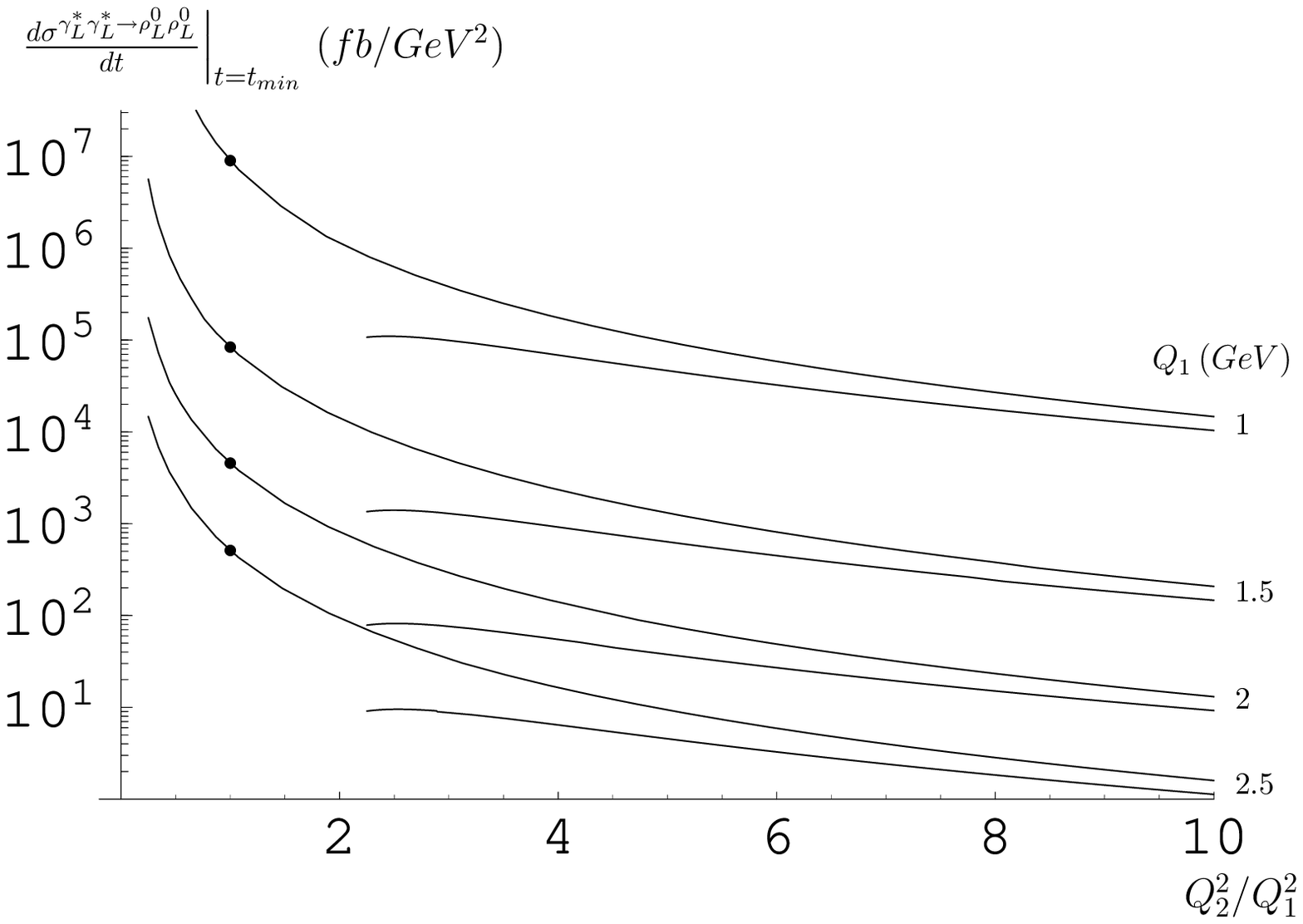}}}
\end{minipage}
\begin{minipage}[t]{105mm}
\epsfxsize=7.7cm{\centerline{\hspace{-7.6cm}\epsfbox{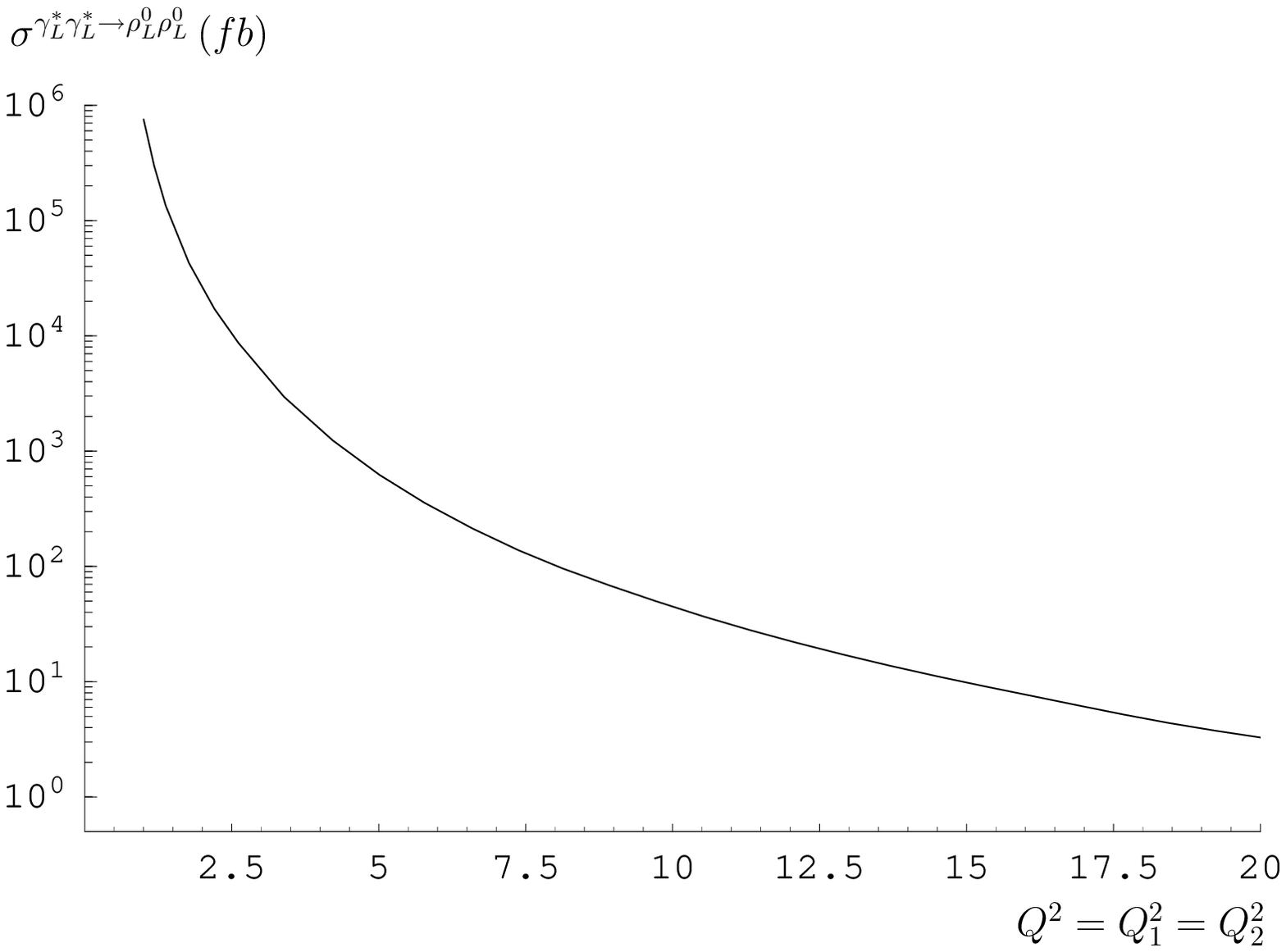}}}
\end{minipage}
\caption{Left: Differential cross-section
  for the process $\gamma^*_L\gamma^*_L \to \rho^0_L\rho^0_L$
at Born order, at the threshold $t=t_{min},$ as a function of  
$Q_2^2/Q_1^2.$
The dots represents the value of the cross-section at the special point
$Q_1=Q_2,$
as given by the analytical formula (\ref{amplitudeR1}). The
asymptotical curves are valid for large $Q_2^2/Q_1^2$ and are given 
by Eq.(\ref{amplitudeparton}). 
~~~Right: The integrated cross-section
  for the process $\gamma^*_L\gamma^*_L \to \rho^0_L\rho^0_L$
at Born order as a function of $Q_1^2=Q_2^2.$}
\label{Figtmin}
\end{figure}
 In Fig.1L we display the differential cross-section 
\beq
\label{crosssection}
\frac{d \sigma^{\gamma^*_L \gamma^*_L \to \rho^0_L  
\rho^0_L}}{dt}=\frac{|{\cal M}|^2}{16
\, \pi \,s^2 }\,
\eq
for vanishing transverse $t-$channel
       momentum, {\it i.e.} $t=t_{min},$ as a
function of the ratio $Q_2^2/Q_1^2.$ Curves are labelled by the values  
of
       $Q_1.$
The dots on the curves represent the values of the cross-section at the  
special point
$Q_1=Q_2,$ which obviously correspond to the analytical formula  
(\ref{amplitudeR1}).

The peculiar limits $R \gg 1$ and $R \ll 1$ are of special physical interest,
since they correspond to the kinematics typical for deep inelastic scattering
on a photon target described through collinear approximation, {\it i.e.} the usual
parton model.
In the limit $R \gg 1$, the amplitude simplifies into
\beq
\label{amplitudeparton}
{\cal M}_{t_{\min}} \sim i s \frac{N_c^2-1}{N_c^2} \, \alpha_s^2 \, \alpha_{em} \,
\alpha(k_1) \, \beta(k_2) \, f_\rho^2 \, \frac{96 \pi^2}{Q_1^2\, Q_2^2}
\left(\frac{\ln R}{R}-\frac{1}{6 R} \right)\,.
\eq
The corresponding asymptotical curves are shown on Fig.1L.

\section{Integrated cross section}

We get the $t-$ dependence of the cross section by computing exactly the 
two dimensionnal integral (\ref{defM}) and then performing the integration
over $z_1$ and $z_2$ numerically in order to get the amplitude 
(\ref{MCalgeneral}). As may be  
anticipated, the cross-section is strongly peaked in the forward  
direction. This does not mean that most particles escape detection since the  
virtual
  photons are not in the direction of the beams.
The differential cross-section shows up sufficient for the
$t-$dependence to be measured up to a few ${\rm GeV^2.}$

Figure 1R shows the integrated over $t$ cross-section as a
function of $Q^2=Q_1^2=Q_2^2.$ The magnitude of the cross-section appears  
to be
sufficient for a detailed study to be performed at the linear collider  
presently
under study. Note that we did not included the virtual photon  
fluxes, which
would amplify the dominance of smaller $Q^2.$  However, triggering  
efficiency
often increases substantially with $Q^2$ \cite{royon}.
At this level of calculation there is no $s$-dependence of the
cross-section. It will appear after taking   into account of BFKL  
evolution \cite{EPSW}.

\section{Conclusion}
This Born order study shows that the process   
$\gamma^*_L(Q_1^2)\gamma^*_L(Q_2^2) \to
\rho_L^0 \rho_L^0$ can be measured  at foreseen $e^+-e^--$colliders for
$Q_1^2, Q_2^2$ up to a few ${\rm GeV^2}.$
  Indeed, a nominal integrated luminosity of $100 \,{\rm fb}^{-1}$ should
yield thousands of events per year, with ${ Q^2 \gtrsim 1}$ GeV$^2$.
  This would open
a new domain of investigation for diffractive processes in which  
practically
all ingredients of the scattering amplitude are under control within a  
pertubative
approach.

  BFKL evolution\cite{EPSW} is expected to give a net and visible  
enhancement of
   the cross section.

\section*{Acknowledgements}

  We thank R.~Enberg for discussions. L.Sz. is a Visiting Fellow of
the Fonds National pour la Recherche Scientifique (Belgium).
Work of L.Sz. is supported by Polish Grant 1 P03B 028 28.



\section*{References}
\begin{thebibliography}{99}

\bibitem{bfkl} E.A.~Kuraev,  L.N.~Lipatov and V.S.~ Fadin,
Phys.~Lett. {\bf B60} (1975) 50-52; 
Sov.~Phys.~JETP {\bf 44} (1976) 443-451;
  Sov.~Phys.~JETP {\bf 45} (1977) 199-204;
Ya.Ya.~Balitskii and L.N.~Lipatov,
Sov.~J.~Nucl.~Phys. {\bf 28} (1978) 822-829.

\bibitem{PSW}
B.~Pire, L.~Szymanowski and S.~Wallon,
   arXiv:hep-ph/0507038, Eur.P.J. (2005) in print

\bibitem{EPSW}
R.~Enberg, B.~Pire, L.~Szymanowski and S.~Wallon, arXiv:hep-ph/0508134;
D.~Y.~Ivanov and A.~Papa, arXiv:hep-ph/0508162.

\bibitem{L3}
P.~Achard {\it et al.}  [L3 Collaboration],
Phys.\ Lett.\ B {\bf 568}, 11 (2003);
Phys.\ Lett.\ B {\bf 597}, 26 (2004) and arXiv:hep-ex 0504016.

\bibitem{apt}
I.~V.~Anikin, B.~Pire and O.~V.~Teryaev,
Phys.\ Rev.\ D {\bf 69}, 014018 (2004) and hep-ph 0506277.

\bibitem{royon}
  M.~Boonekamp, A.~De Roeck, C.~Royon and S.~Wallon,
  Nucl.\ Phys.\ B {\bf 555}, 540 (1999)
  [arXiv:hep-ph/9812523].
\end{thebibliography}
\end{document}